\documentclass[aps,prd,twocolumn,nofootinbib,floatfix]{revtex4-1}
\usepackage{setspace}
\usepackage{amsthm}
\theoremstyle{definition}

\usepackage{graphicx}
\usepackage{dcolumn}
\usepackage{multirow}
\usepackage{subfigure}
\usepackage{times,mathptm}
\usepackage{float}
\usepackage{color}
\usepackage{amsmath,amsfonts}
\usepackage{mathptmx}
\usepackage{mathrsfs}
\usepackage{bbm}
\usepackage{bm}
\usepackage{xfrac}

\newcommand{\beq}{\begin{equation}}
\newcommand{\eeq}{\end{equation}} 
\newcommand{\bea}{\begin{eqnarray}}
\newcommand{\eea}{\end{eqnarray}}

\newcommand{\E}{\mathcal{E}}

\newcommand{\Sc}{S$_\text{c}$}

\renewcommand{\d}{\delta}

\renewcommand{\l}{\lambda}

\newcommand{\T}{{\cal T}}

\renewcommand{\b}{\beta}

\newcommand{\tr}{\text{Tr}}

\newcommand{\vx}{{\vec{x}}}
\newcommand{\vy}{{\vec{y}}}
\newcommand{\vz}{\vec{z}}

\newcommand{\n}{\nu}
\newcommand{\m}{\mu}
\newcommand{\q}{\overline{q}}
\newcommand{\g}{\gamma}

\renewcommand{\th}{\theta}

\newcommand{\dg}{\dagger}
\newcommand{\non}{\nonumber}

\newcommand{\rf}[1]{(\ref{#1})}
\newcommand{\ra}{\rightarrow}
\newcommand{\pa}{\partial}
\renewcommand{\vec}[1]{\bm #1}
\usepackage{ulem}

\begin{document}

\title{Excitations of elementary fermions in gauge Higgs theories} 

\bigskip
\bigskip

\author{Jeff Greensite}
%\singlespacing
\affiliation{Physics and Astronomy Department \\ San Francisco State
University  \\ San Francisco, CA~94132, USA}
\bigskip
\date{\today}
\vspace{60pt}
\begin{abstract}

\singlespacing
 
    Static quark-antiquark states in QCD, at finite quark separation, have a spectrum of metastable states corresponding
to string-like excitations of the gauge field.   In this article I suggest that there may also exist an excitation spectrum of heavy fermions in some gauge Higgs theories deep in the Higgs phase.  In this situation there are no color electric flux tubes connecting quarks with antiquarks.  There may, nonetheless, exist  stable excitations of the bosonic fields surrounding an isolated fermion, below the particle production threshold.  I present numerical evidence indicating the existence of such excitations in an SU(3) gauge Higgs theory, with the scalar field in the fundamental representation of the gauge group.  
   
\end{abstract}

\maketitle

\singlespacing
%\begin{widetext}
%\section{\label{intro}Introduction}

\section{\label{Intro} Introduction}

   It has long been known that static quark-antiquark states in QCD have a spectrum of string-like excitations
of the color electric field joining the quarks.  In a system with light dynamical quarks these excitations are of course
only metastable, due to string breaking, and indeed the light quark-antiquark states themselves have a spectrum of
metastable excitations, lying on linear Regge trajectories.  We expect the same phenomena in the confinement phase
of a gauge Higgs theory, with the scalar field in the fundamental representation of the gauge group.  

   In the Higgs phase of a gauge Higgs theory, however, there are no color electric flux tubes, and therefore no
spectrum of string excitations associated with an isolated fermion.  I will argue in this article, however, that there may
still exist excited states of isolated fermions,  corresponding to a spectrum of excitations of the surrounding
gauge and Higgs fields.   Lattice Monte Carlo evidence for such a possibility is presented below, in an SU(3) gauge Higgs theory with a unimodular scalar field in the fundamental representation of the gauge group.

\section{\label{Sconf} Varieties of confinement}

\begin{figure}[t!]   
\centerline{\includegraphics[scale=0.3]{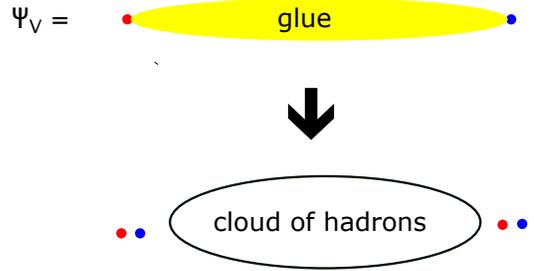}}
\caption{Decay of a state with widely separated quark-antiquark color charges and fractional electric charge into
a set of color neutral hadrons of integer electric charge.  The property of \Sc\ confinement is related to the 
energy of the color charge separated state $\Psi_V(R)$, in the limit of color charge separation $R \ra \infty$.}
\label{psiV}
\end{figure}

   Let us begin with ordinary QCD, and ask the question: what is the binding energy of a hadron, e.g.\ the J/$\psi$?
Of course it is impractical to address this question experimentally.  Any attempt to ``ionize'' a quarkonium state will just result in
more hadrons, rather than a well separated pair of color-charged particles.  Nevertheless, there exist states
in the physical Hilbert space which correspond to precisely that latter situation.   For massive, static quarks, such states
have the form
\beq
        \Psi_V(R) = \q^a(\vx)  V^{ab}(\vx,\vy;U) q^b(\vy) \Psi_0 \ ,
\label{PsiV}
\eeq
where $\q, q$ are quark/antiquark operators transforming in the fundamental representation of the gauge group, $a,b$ are
color indices, and $\Psi_0$ is the vacuum state.  The $V$ operator transforms under a gauge transformation like a Wilson line running between points $\vx$ and $\vy$, and depends {\it only} on the gauge field $U_i$ (we assume throughout a lattice regularization),
and not on the quark fields or any other matter fields.   
In QCD, a state of this kind might represent a quark of electric charge $+2/3$, some long distance away from an antiquark 
of electric charge $-2/3$, with no other electric charge in the region.  Of course such a state would not persist for long, and would soon decay into a set of integer charged
hadrons, as indicated in Fig.\ \ref{psiV}.  But the point is that states with a large separation between a quark-antiquark pair, unscreened by any other matter
fields, do exist in the Hilbert space.  We are interested in how the energy of this subclass of states varies with quark separation.

  Let $E_V(R)$, with $R=|\vx-\vy|$, be the expectation value of the energy of the state $\Psi_V$ above the vacuum energy.  We say that the binding energy of a $q\q$ state 
is infinite, or that QCD has the property of  ``separation-of-charge''  (\Sc) confinement  \cite{Greensite:2017ajx}, iff
\beq
          \lim_{R \ra \infty} E_V(R) = \infty  
\label{Sc}
\eeq
for {\it any} choice of the operator $V(\vx,\vy;U)$, again with  the important restriction that $V$ is a functional of the
gauge field only.  Of course it is always possible to choose a particular $V$ such that this condition is satisfied; an example
is a Wilson line running between $\vx$ and $\vy$, in which case \rf{Sc} holds even in a non-confining theory such as QED.  The \Sc\ condition requires that the condition is satisfied for {\it every}  $V$.

    Now suppose, instead of QCD, we consider a pure gauge theory with only massive, static quarks at points $\vx,\vy$.
Then at any $R$ there is a spectrum of energy eigenstates
\beq 
 \Psi_n(R) = \q^a(\vx)  V_n^{ab}(\vx,\vy;U) q^b(\vy) \Psi_0 \ ,
\label{Pn}
\eeq
which correspond to the ground and excited states of a color electric flux tube, running between points $\vx, \vy$.  Such
a spectrum has in fact been observed in lattice Monte Carlo simulations \cite{Juge:2002br}.\footnote{For a Nambu-Goto string
stretched between two fixed points, the excitation spectrum was derived by Arvis \cite{Arvis:1983fp}.}  Of course states whose excitation energy exceeds the mass of a glueball cannot be energy eigenstates, since they can emit a glueball and fall into a lower energy state. But below this limit the spectrum is stable.  In QCD there are no stable flux tube states for separations $R$ greater than some critical distance, due to string breaking.  Nevertheless, some of the $\Psi_n$ may still exist as metastable states in the theory.  In fact, in QCD, this is exactly the case for the resonances which lie on linear Regge trajectories.  These are not stable
states, of course, but rather correspond to metastable excitations of the color electric flux tube.

   Let us now consider gauge Higgs theories, with the Higgs scalar in the fundamental representation of the gauge group.  As was emphasized recently by Matsuyama and myself in ref.\ \cite{Greensite:2020nhg}, such theories
have at least two distinct phases:  a confinement phase, with the property of \Sc\ confinement defined above, and a Higgs phase in which the \Sc\ confinement property is lost, and the global $Z_N$ center subgroup of the local SU(N) gauge group is spontaneously
broken.   The Higgs phase turns out to be closely analogous to a spin glass phase, and there is a gauge invariant order
parameter for the confinement-to-Higgs transition which is a direct translation, from condensed matter to a gauge
theory context, of the Edwards-Anderson order parameter for spin glass transitions.   The asymptotic spectrum of the
Higgs phase still consists of color singlets, along the lines discussed by Fr\"{o}hlich, Morcio and Strocchi \cite{Frohlich:1981yi} and
`t Hooft \cite{tHooft:1979yoe} (see also Maas et al.\ \cite{Maas:2017xzh}).  We refer to this property as ``color'' (C)  confinement;
it is much weaker than the \Sc\ confinement condition. Thus the Higgs phase is distinguished from the confinement phase both by
symmetry, and by type of confinement.  The two phases are not, however, necessarily separated by some line of non-analyticity in the free energy \cite{Fradkin:1978dv,Osterwalder:1977pc}.  In this sense the transition may be analogous to a Kertesz line \cite{Kertesz}.

\section{\label{psmat} Pseudomatter fields}

   In discussing the spectrum of elementary fermions in the Higgs phase, the concept of a ``pseudomatter'' operator will be crucial.  A pseudomatter operator $\rho^a(\vx;U)$ is a non-local functional of the gauge field which transforms, at point $\vx$, like
a matter field in the {\it fundamental} representation of the gauge group, with the important exception that, like the gauge field itself, it is insensitive to transformations in the global center subgroup of the gauge group. Hence a pseudomatter operator transforms
in the same way under gauge transformations $g(x)$ and $zg(x)$, where $z$ is an element of the global center subgroup. The simplest example of a pseudomatter operator, in an infinite volume, comes from the abelian theory
\bea
            \rho(\vx;A) &=& \exp\left[-i {e\over 4\pi} \int d^3z ~ A_i(\vz) {\pa \over \pa z_i}  {1\over |\vx-\vz|}  \right] \ . \non \\
\label{Dirac}
\eea
Let us consider gauge transformations $g(x) = e^{i\th(x)}$, and we separate out the zero mode $\th(x) = \th_0 + \widetilde{\th}(x)$.
It is easy to verify that under such transformations
\beq
\rho(\vx;g\circ A) = e^{i\widetilde{\th}(x)} \rho(\vx;A) \ .
\eeq
Using this pseudomatter field, one may construct physical states corresponding to an isolated point charge 
\beq
           \Psi' =  \rho^\dg(\vx;A) \psi(\vx) \Psi_0
\eeq
This construction is very well known, and was first introduced by Dirac \cite{Dirac:1955uv}.  Note that while the operator $\rho^\dg(\vx;A) \psi(\vx)$ is invariant under local gauge transformations, it still transforms under global U(1) transformations. This is the hallmark of an operator which can create a physical state associated with a definite isolated charge, given that the ground state
is itself an eigenstate of zero charge.

    Note also that the gauge transformation defined
by $g_C(\vx;A) = \rho^\dg(\vx;A)$ is precisely the transformation to Coulomb gauge, so in that gauge $\Psi' = \psi(\vx)\Psi_0$.
The fact that a gauge choice defines a set of pseudomatter operators is quite general, and is not restricted to the abelian theory.  Let $g_F^{ab}(\vx;U)$ be the transformation to a physical gauge defined by some condition $F(U)=0$ imposed on spacelike links in an SU($N$) gauge theory.  Then we
may always express $g^\dg_F(\vx;U)$ at any point $\vx$ in terms of its eigenvectors (enumerated by the index $n$)
\beq
         g^{\dg an}_F(\vx;U) = u^a_{(n)}(\vx;U)  ~~~,~~~ u^{\dg a}_{(m)}(\vx;U) u^a_{(n)}(\vx;U) = \d_{mn} \ .
\label{gF}
\eeq
Now let $g$ be any infinitesimal gauge transformation.  Then $g_F$ must have the property
\beq
       g^\dg_F(\vx;g \circ U) = g(\vx) g^\dg_F(\vx;U) \ ,
\eeq
which means that
\beq
           u_{(n)}(\vx;g \circ U) = g(\vx) u_{(n)}(\vx;U) \ ,
\eeq
from which we conclude that $u_{(n)}(\vx;U)$ is a pseudomatter field.           
This observation can be turned around: From a set of $N$ orthogonal pseudomatter fields, with orthogonality defined by
\beq
          \sum_{\vx} \rho^{\dg a}_n(\vx;U) \rho^a_m(\vx;U) = \d_{nm} \ ,
\eeq
it is possible to construct another set of pseudomatter fields $u^a_{(n)}(\vx;U)$ which define a gauge choice, i.e.\ a transformation $g_F$ to some physical gauge. This is the logic of the Laplacian gauge introduced by Vink and Wiese in \cite{Vink:1992ys}, and the procedure for constructing the $u_{(n)}(\vx;U)$ from a set of pseudomatter operators $\rho_n(\vx;U)$ is outlined in
that reference.  

   The eigenstates $\zeta_n(\vx;U)$ of the covariant lattice Laplacian operator
\beq
           (-D_i D_i)^{ab}_{\vx \vy} \zeta^b_n(\vy;U) = \l_n \zeta^a_n(\vx;U)  \ ,
\eeq
where
\bea
 \lefteqn{(-D_i D_i)^{ab}_{\vx \vy} = } \non \\
    &=& \sum_{k=1}^3 \left[2 \d^{ab} \d_{\vx \vy} - U_k^{ab}(\vx) \d_{\vy,\vx+\hat{k}}  - U_k^{\dg ab}(\vx-\hat{k}) \d_{\vy,\vx-\hat{k}}   \right] \ , \non \\  
\label{Laplacian}
\eea
are all examples of pseudomatter fields, and will be especially important here. Once again, these fields transform like matter in the fundamental representation of the gauge group, apart from their invariance under the global $Z_N$ subgroup of the SU(N) gauge group.  The ``Laplacian Landau gauge'' introduced by Vink and Wiese \cite{Vink:1992ys} made use of the low-lying eigenstates of the Laplacian operator in four Euclidean dimensions.  In the next section we will be concerned
with the low-lying eigenstates of the three dimensional lattice Laplacian operator \rf{Laplacian}, defined at fixed time on a $D=4$ dimensional lattice.\footnote{These eigenstates are computed numerically via the Arnoldi algorithm, as implemented in
the ARPACK software package (https://www.caam.rice.edu/software/ARPACK/).}   

   It was shown in \cite{Greensite:2020nhg} that in the spin glass (i.e.\ Higgs) phase of the gauge Higgs theory, it is always possible to find a physical gauge defined by $F(U)=0$ such that $\langle \phi \rangle$ is
non-zero, i.e.\ 
\beq
        \langle g_F(\vx;U) \phi(\vx) \rangle \ne 0 \ .
\label{zbreak}
\eeq
 A corollary is the loss of \Sc\ confinement in the Higgs phase.  If $g_F(\vx;U)$ is the gauge transformation
to a gauge in which $\langle \phi \rangle$ is non-zero (and $g_F$, as just pointed out, can always be decomposed into a
set of pseudomatter fields), then one may choose
\beq
          V^{ab}(\vx,\vy;U) = g_F^{\dg ac}(\vx;U) g_F^{cb}(\vy;U) \ ,
\label{VF}
\eeq
and show that  $E_V(R)$ has a finite limit at $R\ra \infty$.  Conversely, in the phase of unbroken global $Z_N$ gauge symmetry, 
and assuming the absence of a massless phase, we must have $E_V(R) \ra \infty$, i.e.\ \Sc\ confinement, in the same limit.  For details, cf.\  \cite{Greensite:2020nhg}.
Physical quark-antiquark states in the confined phase, with finite energy in the $R\ra \infty$ limit, are 
created by operators such as
\beq
       Q(\vx,\vy) =  [\q^a(\vx)  \phi^a(\vx)] \times [\phi^{\dg b}(\vy) q^b(\vy)] \ ,
\eeq
which can be thought of as creating two color neutral quark-scalar bound states.

\subsection{Pseudomatter in finite volumes}

    One cannot create a single electric charge in a finite volume with periodic boundary conditions.  The reason that the construction \rf{Dirac} doesn't work in a finite volume, in ordinary QED, is that the equation $-\nabla^2 D(\vz) = \d^3(\vx-\vz)$ is not soluble in a finite periodic volume.  Instead one must create $\pm$ charges in pairs, as in \rf{PsiV}, with
\bea
         V(\vx,\vy;A) &=& \exp\left[-i e \int d^3z ~ A_i(\vz) {\pa \over \pa z_i} D(\vz) \right]  \non \\
            -\nabla^2 D(\vz) &=&  \d^3(\vz-\vx) - \d^3(\vz-\vy) \ .
\eea
Likewise, in the non-abelian case, the eigenstates $\zeta_n(\vx;U)$ are determined only up to a global gauge-invariant phase.
Unless these operators 
occur in pairs such that the global phases cancel, i.e.\ $\zeta^a_n(\vx;U) \zeta^{\dg b}_n(\vy;U)$, they will vanish in expectation values due to wild fluctuations in the global phase.

   In general, in a finite volume, we consider in the Higgs phase operators of the form
\beq
         V^{ab}(\vx,\vy;U) = \sum_n c_n \rho^a_n(\vx;U) \rho^{\dg b}_n(\vy;U) \ ,
\eeq
and consider taking the $R=|\vx-\vy| \ra \infty$ limit (along with the infinite volume limit).  
Then instead of \rf{zbreak}, the criterion for spontaneous breaking of
global $Z_N$ gauge symmetry is the existence of a finite limit in the correlator
\beq
         \lim_{R\ra \infty} |\langle \phi^{\dg a}(\vx) V^{ab}(\vx,\vy;U) \phi^b(\vy) \rangle| > 0 \ ,
\eeq
for some $V$.  In the Higgs phase, there will always exist a transformation $g_F(\vx;U)$ to some $F$-gauge,  such that this criterion is satisfied by the $V$ operator in \rf{VF}.\footnote{There is an alternative approach to estimating the left hand side of \rf{zbreak} in some $F$-gauge, which is generally employed in computer simulations, via computation of the quantity
\beq
          {1\over V} \left\langle \sum_{t=1}^{L_t} \left| \sum_{\vx} g_F(\vx;U) \phi(\vx,t) \right| \right\rangle \non
\eeq
on a $V=L^3 \times L_t$ lattice volume, and extrapolation to infinite spatial volume.  The modulus of the sum over $\vx$ is used to eliminate the ambiguity with respect to any remnant global symmetry transformations in the $F$-gauge.}

\section{\label{exite} Excitations of fermions}

   I now put forward the conjecture that just as there is a set of metastable states \rf{Pn} in the confined phase at fixed $R$,
so there is also a spectrum of excitations of a static fermion-antifermion system in the Higgs phase, at least for some gauge Higgs theories, with a finite energy above the ground state out to $R\ra \infty$.  The term ``ground state'' now refers not to the vacuum, but to the minimal energy state containing a static fermion-antifermion pair.   It is supposed that this gap in energy is too small to be explained simply by the presence of additional vector or Higgs bosons. I support this conjecture with an example.  The model is SU(3) lattice gauge theory with a standard Wilson action and a unimodular ($\phi^\dg(x) \phi(x) = 1$) Higgs field in the fundamental representation:
\bea
     S  &=& - {\beta \over 3} \sum_{plaq}  \mbox{ReTr}[U_\m(x)U_\n(x+\hat{\m})U_\m^\dg(x+\hat{\n}) U^\dg_\n(x)]  \non \\
         & &       - \gamma \sum_{x,\m}  \mbox{Re}[\phi^\dg(x) U_\m(x) \phi(x+\widehat{\m})] \ . 
\label{Sgh}
\eea
The methods of ref.\ \cite{Greensite:2020nhg} can be used to determine the transition between the \Sc\ confining and the spin glass (Higgs) phases.  In this article I will work at the Wilson coupling $\b=5.5$ and a variety of $\gamma$.  At this $\b$ value
the extrapolation method of \cite{Greensite:2020nhg} yields an estimate of $\g = 1.35(3)$ at the transition.  

\begin{figure*}[t!]
\subfigure[~]  % caption for subfigure a
{   
 \label{olap1}
 \includegraphics[scale=0.6]{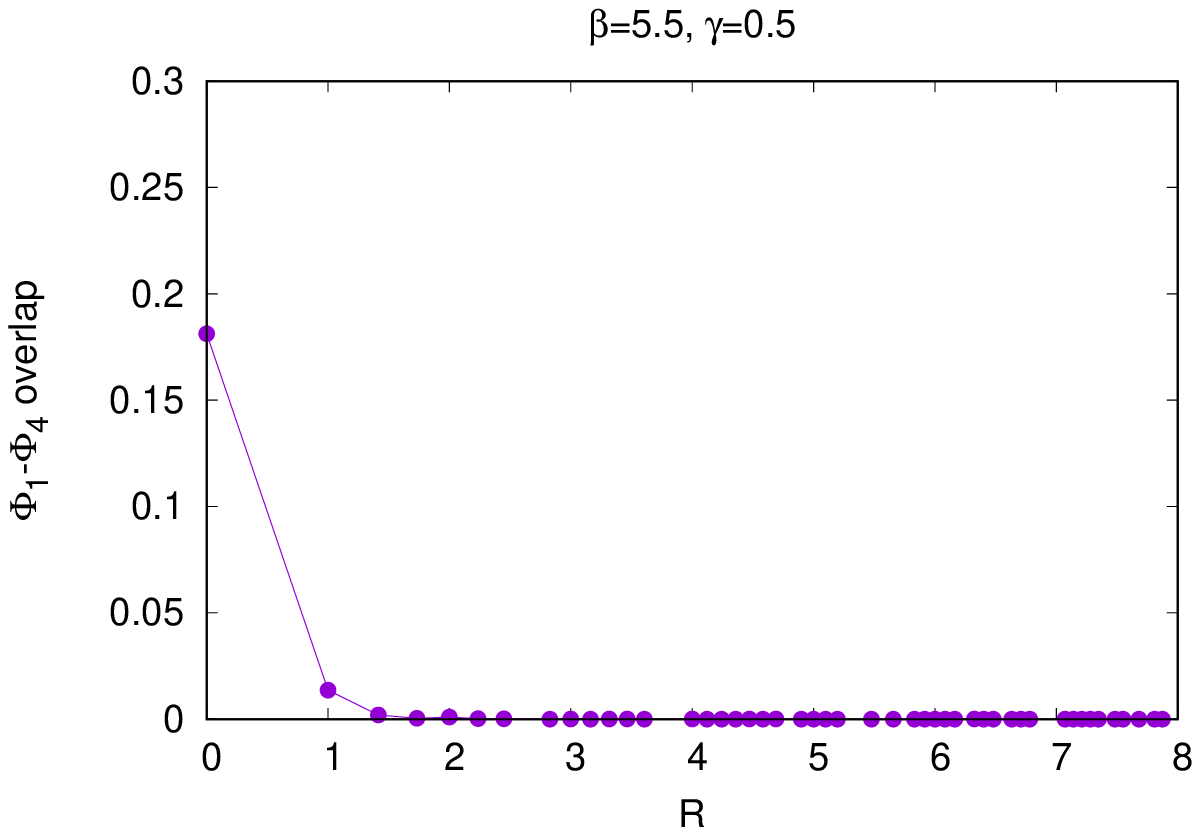}
}
\subfigure[~]  % caption for subfigure a
{   
 \label{olap2}
 \includegraphics[scale=0.6]{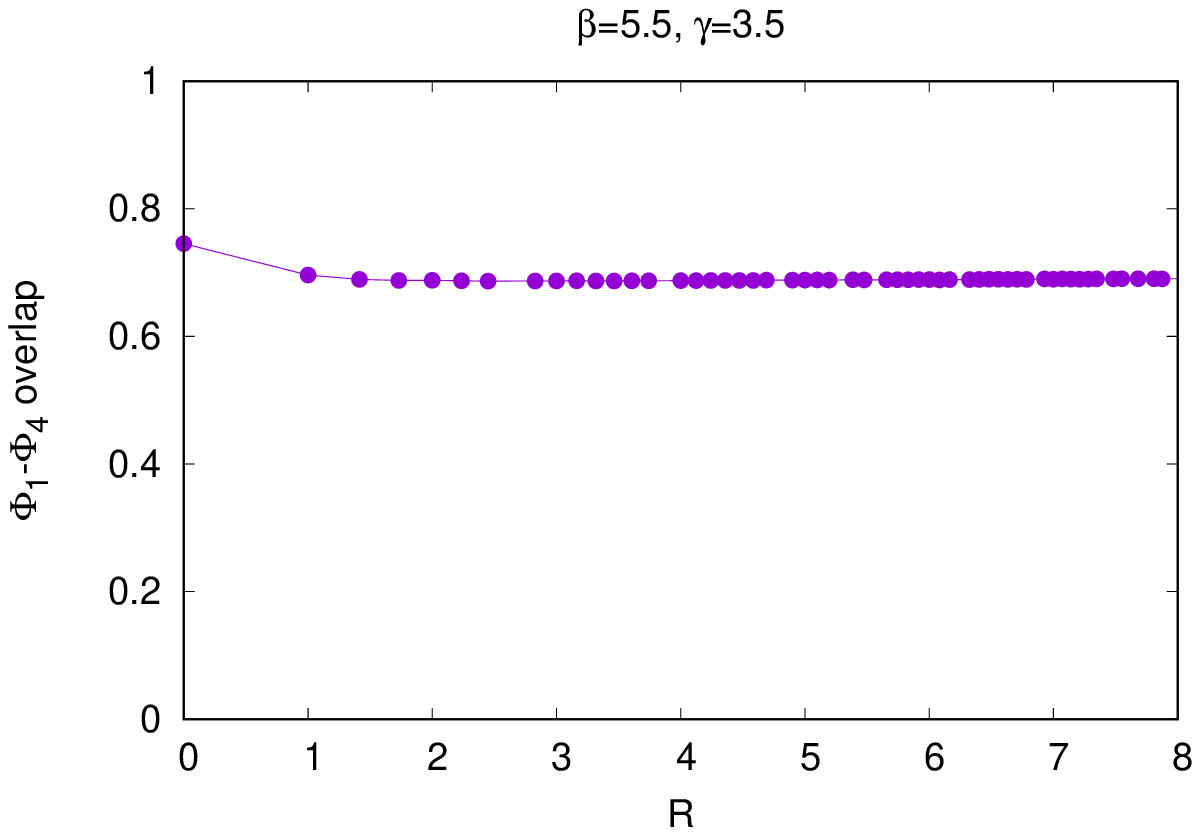}
}
\subfigure[~]  % caption for subfigure a
{   
 \label{EV05}
 \includegraphics[scale=0.6]{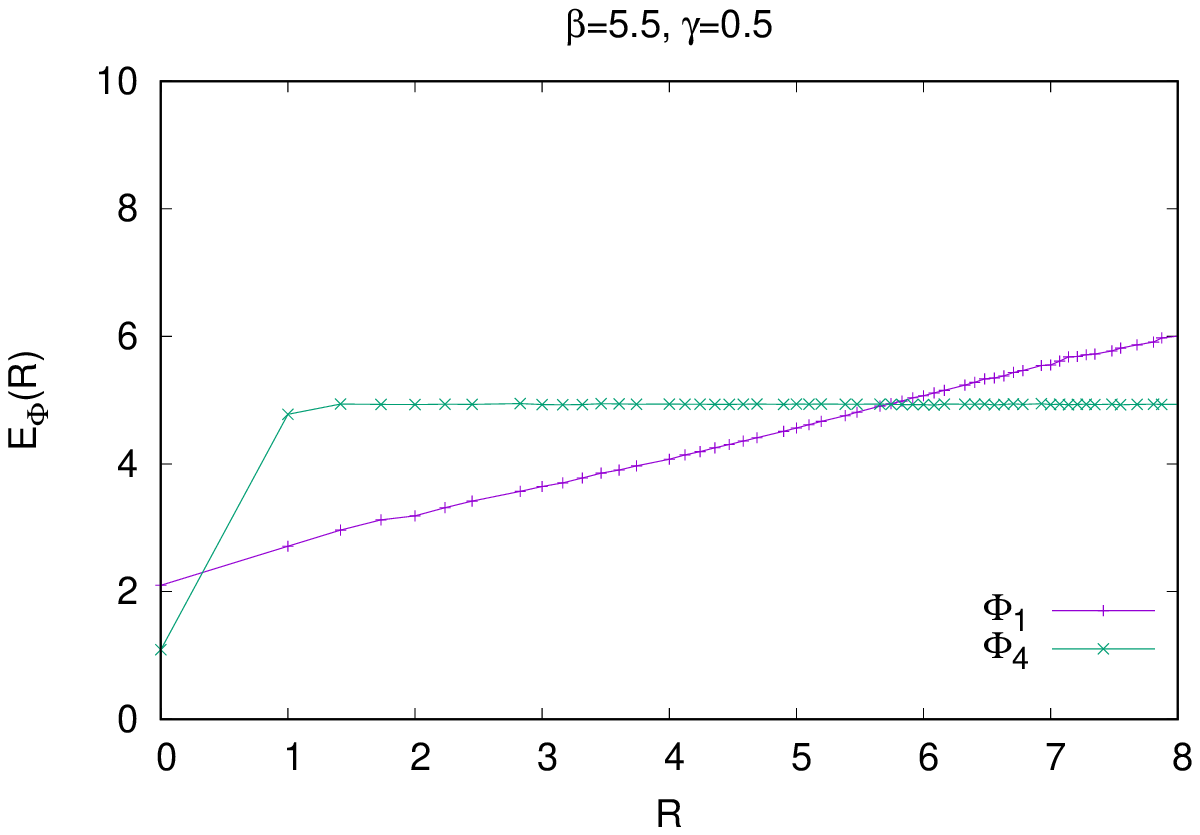}
}
\subfigure[~]  % caption for subfigure a
{   
 \label{EV35}
 \includegraphics[scale=0.6]{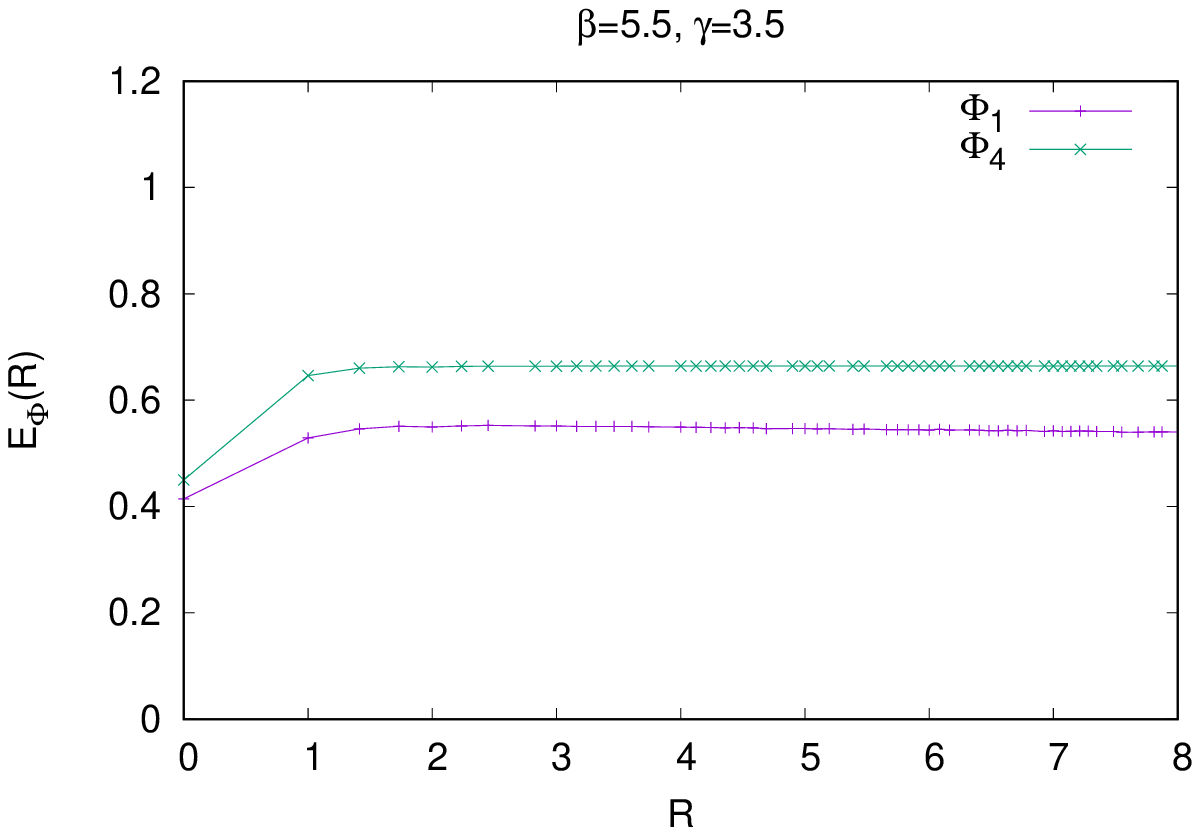}
}
\caption{Contrasting properties of pseudomatter states in the confinement and Higgs phases of an SU(3) gauge Higgs theory. (a) Overlap vs.\ $R$ of normalized fermion-antifermion states using pseudomatter ($\Phi_1$) and the Higgs field
($\Phi_4$) states in the confined phase, at $\b=5.5, \g=0.5$. (b) Same as subfigure (a), but in the Higgs phase at 
$\b=5.5, \g=3.5$. (c) Energy expectation value $E^\Phi(R)$ vs.\ separation $R$ of the $\Phi_1$ and $\Phi_4$ states in the confined phase, $\b=5.5, \g=0.5$. (d) Same as subfigure (c), but in the Higgs phase at $\b=5.5, \g=3.5$.}
\label{Psidata}
\end{figure*}
Now consider, at each $R=|\vx-\vy|$, the following set of four (in general 
non-orthogonal) states:
\beq
         \Phi_n(R) = Q_n(R) \Psi_0 \ ,
\label{Phistates}
\eeq
where, for $n=1,2,3$
\beq
         Q_n(R) = [\q^a(\vx) \zeta_n^a(\vx;U)] ~\times~  [\zeta_n^{\dg b}(\vy;U) q^b(\vy)] \ ,
\eeq
and
\beq
          Q_4(R) = [\q^a(\vx) \phi^a(\vx)] ~\times ~  [\phi^{\dg b}(\vy) q^b(\vy)]   \ ,
\eeq
where the $\zeta_n(\vx;U)$ are pseudomatter operators corresponding to eigenstates of the lattice Laplacian with the three largest eigenvalues $\l_n$.  Of course the $\Phi_n$ states all have the form \rf{PsiV}, with
\bea
V^{ab}_n(\vx,\vy;U) &=&   \zeta_n^a(\vx;U) \zeta_n^{\dg b}(\vy;U)  ~~(n=1,2,3) \non \\
V^{ab}_4(\vx,\vy;\phi) &=&    \phi^a(\vx) \phi^{\dg b}(\vy) \ .
\eea
As already mentioned, the Higgs and confinement phases are distinguished by the 
spontaneous breaking of the global $Z_3$ center subgroup of SU(3) gauge symmetry in the Higgs phase, and what this implies is that in the unbroken, \Sc\ confinement phase, $\Phi_4(R)$ is orthogonal to the other three states in the $R\ra \infty$ limit.  The reason is that the operator $\q^a(\vx) \phi^a(\vx)$ is invariant under all gauge transformations, while $\q^a(\vx) \zeta_n^a(\vx;U)$ transforms under the global $Z_N$ subgroup, since $\q$ transforms under this symmetry, while $\zeta_n^a(\vx;U)$ does not.  This  implies the orthogonality just stated, providing the vacuum itself is invariant under the
global $Z_3$ gauge symmetry. 

   In order to compute energy expectation values $E^\Phi_n(R)$ corresponding to the $\Phi_n(R)$ states, we begin from
the Euclidean time identity
\beq
          \langle Q_m^\dg(R,t) Q_n(R,0) \rangle = \langle \Phi_m(R)| e^{-(H-\E_0)t} | \Phi_n(R) \rangle \ ,
\eeq
where $\E_0$ is the vacuum energy, and $Q(R,t)$ indicates that the operator $Q(R)$ is evaluated at time $t$ .  Then we see that the energy of state $\Phi_n(R)$ above the vacuum energy is given by
\beq
           E^\Phi_n(R) = -\left[ {d\over dt} \log  \langle Q_n^\dg(R,t) Q_n(R,0) \rangle \right]_{t=0} \ ,
\eeq
and the appropriately normalized overlap of states $\Phi_m,\Phi_n$ is
\bea
   o_{mn}(R) &=&  { \langle \Phi_m | \Phi_n \rangle \over \sqrt{\langle \Phi_m | \Phi_m\rangle\langle \Phi_n | \Phi_n \rangle}} \non \\
        &=& { \langle Q^\dg_m(R,0) Q_n(R,0) \rangle \over \{ \langle Q^\dg_m(R,0) Q_m(R,0) \rangle
                                                                                              \langle Q^\dg_n(R,0) Q_n(R,0) \rangle \}^{1/2} } \ .\non \\
\label{o_mn}
\eea
This may be generalized.  We define
\beq
           E^\Phi_n(R,T) = -\left[ {d\over dt} \log  \langle Q_n^\dg(R,t) Q_n(R,0) \rangle \right]_{t=T} \ ,
\eeq
and
\beq
          o_{mn}(R,T) =  { \langle Q^\dg_m(R,T) Q_n(R,0) \rangle \over \{  \langle Q^\dg_m(R,T) Q_m(R,0) \rangle
                                                                                              \langle Q^\dg_n(R,T) Q_n(R,0) \rangle \}^{1/2} }  \ .
\eeq                                                                                                                                                                                          
These can be interpreted as the energies and the overlaps of states obtained by evolving the $\Phi_n$ for a Euclidean time interval $T/2$, i.e.\ $\Phi_n(R,T/2) = \exp[-HT/2] \Phi_n(R)$, followed by normalization.

     With discretized time on a hypercubic lattice, the logarithmic time derivative must be replaced by the corresponding lattice
expression
\beq
E^\Phi_n(R,T) = -  \log \left[ { \langle Q_n^\dg(R,T) Q_n(R,0) \rangle \over \langle Q_n^\dg(R,T-1) Q_n(R,0) \rangle} \right] \ .
\eeq
For $T$ an odd integer, this is interpreted as the energy expectation value (minus the vacuum energy) 
of a state evolved for $(T-1)/2$  units of Euclidean time.  The $Q^\dg Q$ correlators are computed on the lattice as follows: Define a timelike Wilson line
\beq
          P(\vx,t,T) = U_0(\vx,t) U_0(\vx,t+1)...U_0(\vx,T-1) \ .
\eeq
Then, after integrating out the static fermions, and discarding, since we are only interested in the energy due to the dynamical fields, an irrelevant quark mass (hopping parameter) factor, 
\bea
       & & \langle Q^\dg(R,T) Q(R,0) \rangle \non \\
       & & = \langle \tr[V^\dg_i(\vx,\vy,U(t+T)) P^\dg(\vx,t,T) V_j(\vx,\vy;U(t)) P(\vy,t,T)]  \rangle \ . \non \\
\eea
In the numerical calculation of this quantity we average over all $\vx,\vy$ with fixed $R=|\vx-\vy|$.

    The overlap $o_{41}(R)$ between the state $\Phi_4(R)$ constructed with the Higgs field, and the state $\Phi_1(R)$
built with the $\zeta_1$ pseudomatter field, is displayed in Fig.\ \ref{olap1} in the confined phase, at ${\b=5.5, ~ \g=0.5}$.
We see that this overlap tends rapidly to zero as $R \ra \infty$, as required by the invariance of the vacuum, in the
confined phase, under global $Z_3$ gauge transformations.  This global subgroup of the gauge symmetry is broken in the Higgs phase, so that $\Phi_1(R)$ and $\Phi_4(R)$ are not necessarily orthogonal in the $R\ra \infty$ limit.  That is what we see in Fig.\ \ref{olap2}, with data obtained in the Higgs phase, at $\b=5.5, ~ \g=3.5$, where the overlap between these states is quite large. 
It is also found in the confinement phase, in Fig.\ \ref{EV05}, that the energy of the quark-pseudomatter state $E^\Phi_1(R)$
rises linearly with $R$, consistent with \Sc\ confinement.  The energy of $E^\Phi_4(R)$ is almost $R$ independent at $R>1$, 
which reflects the fact that $\Phi_4(R)$ consists of a non-interacting pair of color singlet (quark-Higgs) objects. 
In the Higgs phase the energies of both the $\Phi_1$ and $\Phi_4$ states are nearly $R$-independent, as seen in Fig.\ \ref{EV35}.\footnote{It should be noted that the property of \Sc\ confinement in the confinement phase implies that the energies $E_\Phi(R)$
of states $\Phi_{1-3}(R)$ diverge to infinity as $R \ra \infty$.  But it is not necessarily true that the energies of these particular
states have a finite limit as $R \ra \infty$ {\it everywhere} in the Higgs phase, although this finite limit is in fact seen for $\b=5.5$ at
$\g > 1.4$.  While there must always exist, everywhere in the Higgs phase, finite energy states corresponding to 
isolated (i.e.\ $R \ra \infty$) fermions, these need not be the $n=1,2,3$ states listed in \rf{Phistates}, which correspond to a particular choice of the $V$ operator.  For a further discussion of this point, cf.\ \cite{Greensite:2020nhg}.}

   Now let $\tau=\exp(-H)$ be the operator corresponding to the lattice transfer matrix, and we would like to calculate
the eigenstates and eigenvalues of this operator in the Higgs phase, in the four dimensional subspace of Hilbert space
spanned by the non-orthogonal set of states $\{ \Phi_n \}$.  We define the $4\times 4$ matrices
$[\T], [O]$ whose matrix elements are
\bea
          [\T]_{mn} &=& \langle \Phi_m | e^{-(H-\E_0)} | \Phi_n \rangle  \non \\
                          &=& \langle Q_m^\dg(R,1) Q_n(R,0) \rangle \non \\
          \left[ O \right]_{mn} &=& o_{mn} 
\eea 
respectively.   The eigenstates and eigenvalues of $\T=\tau e^{\E_0}$ in the subspace are obtained by solving the generalized eigenvalue problem
\beq
         [\T] \vec{\upsilon}^{(n)} = \lambda_n [O] \vec{\upsilon}^{(n)} \ ,
\eeq
and we have energies above the vacuum energy $\E_0$,  given by
\beq
         E_n(R) = -\log(\lambda_n) \ ,
\eeq
and ordered such that $E_n$ increases with $n$, corresponding to eigenstates in the subspace 
\beq
          \Psi_n(R) = \sum_{i=1}^4 \upsilon^{(n)}_i \Phi_i(R) \ .
\eeq
Likewise we consider evolving the states $\Psi_n$ in Euclidean time
\bea
         \T_{nn}(R,T) &=& \langle \Psi_n | e^{-(H-\E_0)T} | \Psi_n \rangle  \non \\
                                 &=& \upsilon^{*(n)}_i \langle \Phi_i | e^{-(H-\E_0)T}  | \Phi_j \rangle \upsilon^{(n)}_j   \non \\
                                 &=& \upsilon^{*(n)}_i  \langle Q_i^\dg(R,T) Q_j(R,0) \rangle \upsilon^{(n)}_j  
\label{TT}
\eea
 and compute
\beq
           E_n(R,T) =  - \log \left[{\T_{nn}(R,T) \over \T_{nn}(R,T-1) }\right] \ .
\label{ERT}
\eeq
This can be regarded (for $T$ an odd integer) as the energy expectation value of state $\Psi_n(R)$ which has evolved for 
${(T-1)/2}$ units of Euclidean time.   Of course \rf{TT} generalizes to off-diagonal elements $\T_{mn}$ in an obvious way.

There are several possibilities, for each $\Psi_n$:
\begin{enumerate}
\item $\Psi_n(R)$ is an exact eigenstate of the transfer matrix in the full Hilbert space.  Then $E_n(R) = E_n(R,T)$ is time independent.  This situation is rather unlikely.
\item $\Psi_n(R)$ has a substantial overlap with the true ground state, and therefore evolves steadily, in Euclidean time, towards that ground state. Then $E_n(R,T)$ drops rapidly to the lowest possible energy of the static quark-antiquark system with increasing $T$.
\item $\Psi_n(R)$ has very little overlap with the ground state, and rapidly evolves in Euclidean time to a stable or metastable excited state.   Then $E_n(R,T)$ converges to a value which is almost constant, over some range of Euclidean time, above the ground state energy.  This is the interesting situation.  
\end{enumerate}

\begin{figure}[t!]   
\centerline{\includegraphics[scale=0.7]{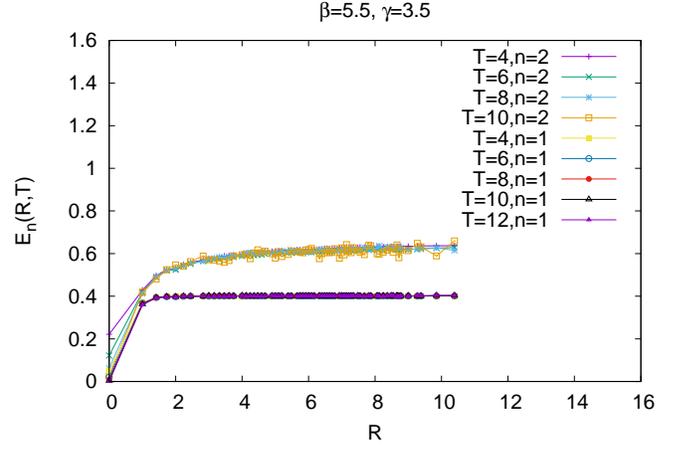}}
\caption{Energies $E_n(R,T)$, defined in \rf{ERT}, of states $\Psi_1, \Psi_2$ after evolution for a period of $(T-1)/2$ units of 
Euclidean time, at lattice couplings $\b=5.5, \g=3.5$.  Note the energy gap, which persists out to the largest $T$ values shown, of $E_2(R,T)-E_1(R,T) \approx 0.2$ in lattice units.}
\label{gap}
\end{figure}

\begin{figure}[t!]
\subfigure[~]  % caption for subfigure a
{   
 \label{EVn1}
 \includegraphics[scale=0.7]{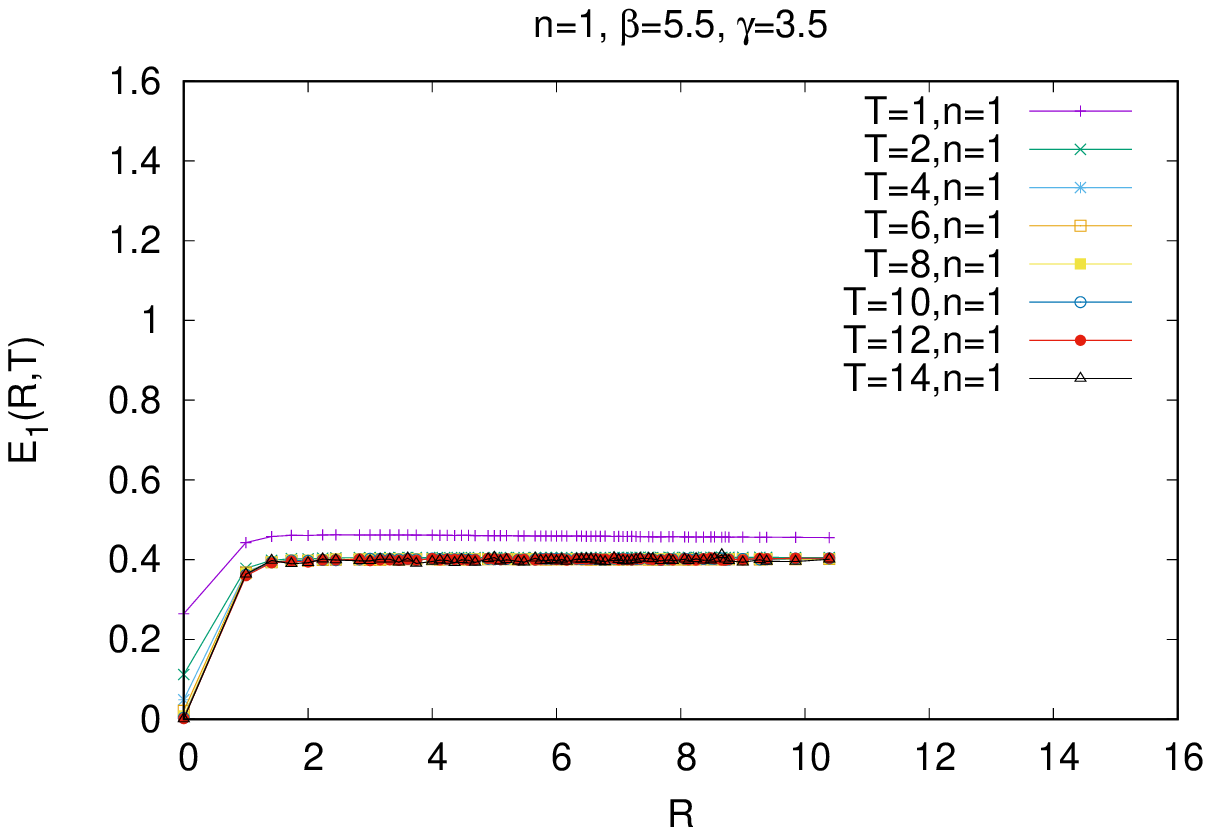}
}
\subfigure[~]  % caption for subfigure a
{   
 \label{EVn2}
 \includegraphics[scale=0.7]{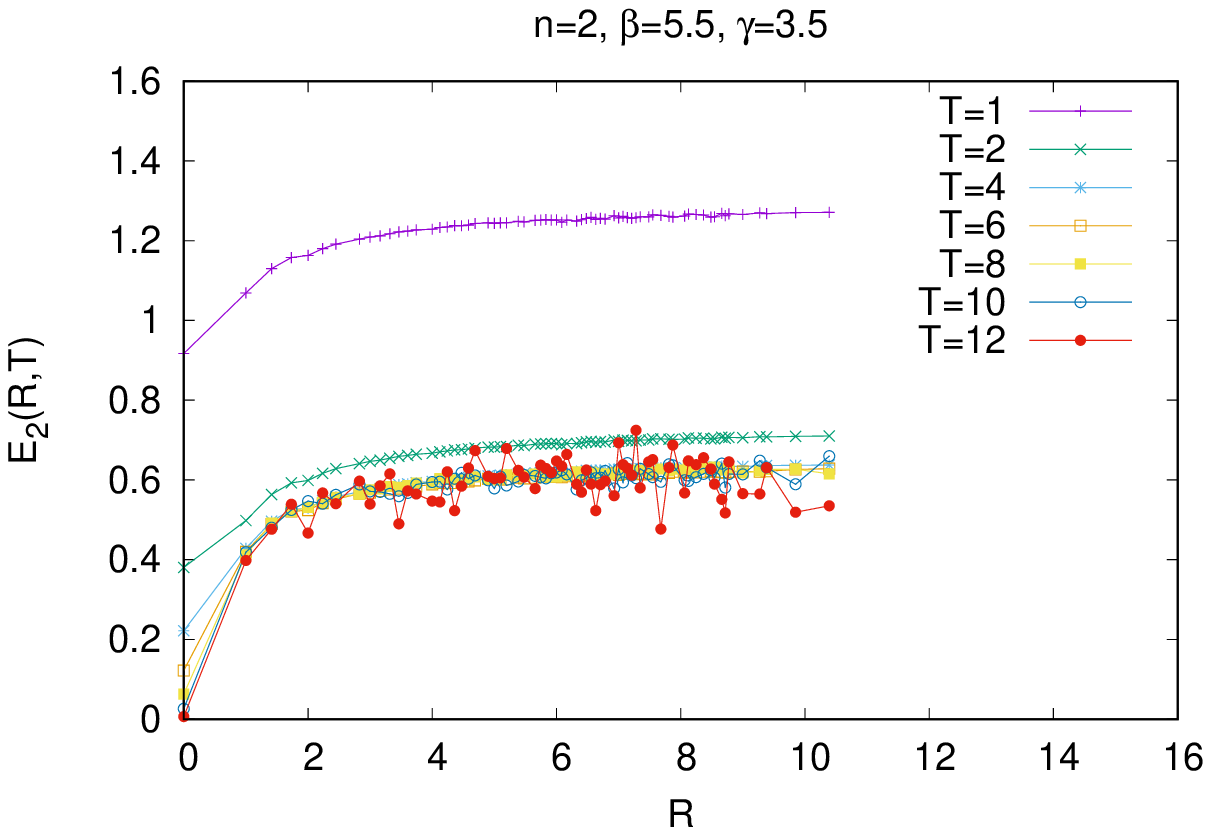}
}
\label{EVn}
\caption{Energies (a) $E_1(R,T)$, and (b) $E_2(R,T)$ at $\b=5.5, \g=3.5$, showing the convergence from $T=1$ to $T=14$ ($n=1$) and $T=12$ ($n=2$).  Neither $\Psi_1$ nor $\Psi_2$ is an energy eigenstate, but both appear to rapidly converge
towards different eigenstates after a short evolution in Euclidean time.}
\end{figure}

   Figure \ref{gap} displays energies $E_n(R,T)$  for $n=1,2$ corresponding to Euclidean time evolution of states $\Psi_{1,2}$, with $T$ ranging from 4 to 12 for the $n=1$ state, and  4 to 10 for the $n=2$ state, with (off-axis) charge separations $R = |\vx-\vy| \le 10$, having components $|x_i-y_i| \le 6$ in the three spatial directions.  The energies are computed in the Higgs phase at $\b=5.5, \g=3.5$, on a lattice volume of $14^3 \times 32$.  Data is obtained from 220 lattices separated by 100 Monte Carlo update sweeps.
For both $n=1,2$ there seems to be a rapid convergence with increasing $T$ to the ground ($n=1$) and an excited ($n=2$) state energy, respectively,  separated by an energy gap of $\approx 0.2$ in units of inverse lattice spacing.  Thus the third possibility, of the three just enumerated, appears to be realized in this theory for the $n=2$ state.  That is the main result, offered in support of
the conjecture that there might exist excited states of non-composite fermions in some gauge Higgs theories.  Data for $E_n(R,T)$ for the $n=3,4$ states is very noisy, and no conclusions can be drawn about them at this stage.   
For $n=2$, the data becomes rather noisy
for $T>10$; nevertheless the data at $T=12$ simply fluctuates around the value obtained at lower $T$.  In Figs.\ \ref{EVn1} and
\ref{EVn2} we
display separately the data for $E_1(R,T)$ and $E_2(R,T)$, including the data at $T=1,2$, and (for $n=2$) the noisy data
at  $T=12$.  Note that while the
$\Psi_{1,2}$ are clearly not energy eigenstates, they converge rapidly in Euclidean time to stable states already at $T=4$. 

\begin{figure}[h!]   
\centerline{\includegraphics[scale=0.7]{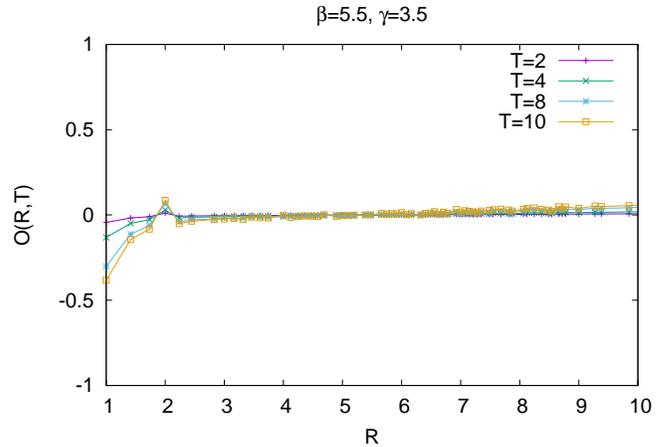}}
\caption{The overlap $\mathcal{O}(R,T)$ between states $\Psi_1(R)$ and $\Psi_2(R)$ which are evolved (and then normalized) for $T/2$ units of Euclidean time.  $\mathcal{O}(R,T)=0$ by construction at $T=0,1$, but the Euclidean time-evolved states are seen to remain approximately orthogonal for $R>2$ at larger $T$. }
\label{ORT}
\end{figure}

    Of course the $\Psi_{1,2}$ states are orthogonal by construction.  But in principle this orthogonality need not persist under Euclidean time (as opposed to real time) evolution, beyond $T=1$.  However, the rapid convergence of $\Psi_{1,2}$ to states with differing energies implies the near-orthogonality of the two states under Euclidean time evolution.  In fact the overlap corresponding to off-diagonal matrix elements
\bea
            \mathcal{O}(R,T) &=&  { \langle \Psi_1 | e^{-HT} | \Psi_2 \rangle \over \sqrt{ \langle \Psi_1 | e^{-HT}  | \Psi_1 \rangle
                       \langle \Psi_2 | e^{-HT}  | \Psi_2 \rangle } } \non \\
                       &=& { \T_{12}(R,T) \over \sqrt{\T_{11}(R,T) \T_{22}(R,T)} }                   
\eea
can be calculated for any $R,T$.  This has the interpretation of an overlap between states obtained from $\Psi_{1,2}$ evolved
for $T/2$ units of Euclidean time, and then normalized.  The result, for $T=2,4,8,10$ (again at $\b=5.5,~\g=3.5$) is shown in Fig.\ \ref{ORT}.  It is clear that the states obtained from evolving $\Psi_1, \Psi_2$ in Euclidean time are very nearly orthogonal, as we had already deduced.
 
   However, there is still the possibility that the energy gap seen in Fig.\ \ref{gap} is not really due to an excitation of the
gauge field surrounding the static fermions, but is rather due to some low momentum particle excitation, e.g.\ a massive vector
boson, or some other particle state.   This would be the case in ordinary QED, where any excited state of a static dipole simply consists of the dipole field plus photons.   In order to rule this out, it is necessary to compare the excitation energy of an excited
state with the mass of the vector boson in the Higgs phase. In this connection it is important to observe that  the particle spectrum of an SU(3) gauge Higgs theory is not necessarily the spectrum that might be expected perturbatively, for reasons that have been discussed at length by Maas et al.\ \cite{Maas:2017xzh}.
Briefly, if one follows the approach of Fr\"{o}hlich, Morcio and Strocchi \cite{Frohlich:1981yi} and
't Hooft \cite{tHooft:1979yoe}, reasoning that particles in the asymptotic spectrum are created by local gauge invariant 
operators or, more precisely, that they show up as poles in the correlation functions of such local operators, then the correspondence between the perturbative and the actual spectrum in the electroweak theory is to some extent coincidental, a
consequence of the approximate SU(2) custodial symmetry, and does not extend to higher gauge groups.

\begin{figure}[h!]   
\centerline{\includegraphics[scale=0.7]{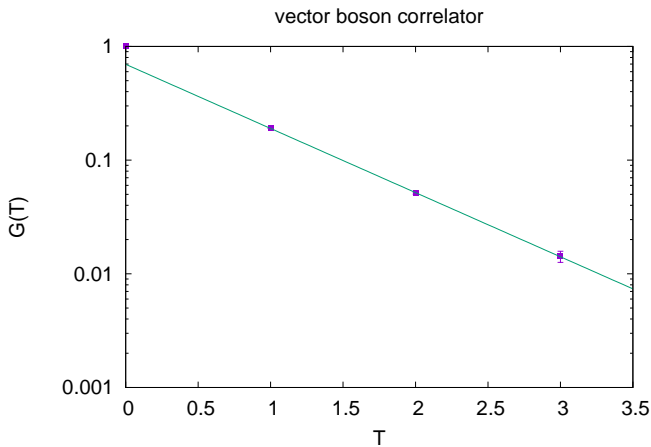}}
\caption{The Euclidean time correlator $G(T)$ (see eq.\ \rf{GT}), shown on a log scale, associated with a zero-momentum vector meson state.  The couplings are $\b=5.5, \g=3.5$. The straight line is a best fit through the three data points at $T>0$, and yields an estimate of the vector boson mass of 1.30(1) in lattice units. }
\label{gmass}
\end{figure}

The spectrum of an SU(3) gauge Higgs theory in the Higgs phase was determined by lattice simulations in \cite{Maas:2018xxu}, in a theory with somewhat different lattice couplings, and with a fluctuating (rather than unimodular) scalar field in the fundamental representation.  It was found that the lightest state was a $1^{--}$ vector meson, whose mass could also be determined analytically
from correlators of the gauge-invariant operator
\beq
           \phi^\dg(\vx) U_k(\vx) \phi(\vx + \hat{k}) \ .
\label{vecbos}
\eeq
Assuming that such an operator also creates the lightest state in the version of SU(3) gauge Higgs theory under consideration here, I have estimated the mass of this lightest state via the standard procedure of projecting, at fixed time, to the zero momentum component
\beq
            Q_k(t) = \sum_{\vx}  \phi^\dg(\vx,t) U_k(\vx,t) \phi(\vx + \hat{k},t) \ ,
\eeq
and then computing the Euclidean time correlation function of the zero momentum operators
\beq
       G(T) = {\langle  \sum_{k=1}^3 Q^\dg_k(T) Q_k(0) \rangle \over \langle \sum_{k=1}^3 Q^\dg_k(0) Q_k(0) \rangle} \ .
\label{GT}
\eeq
The data found on a $14^4$ lattice volume, at $\b=5.5, \g=3.5$, is plotted in Fig.\ \ref{gmass}, and corresponds to a vector
boson mass of 1.30(1) in lattice units, which is more than six times larger than the energy gap of approximately 0.2 in lattice units seen in Fig.\ \ref{gap}.    The conclusion is that the energy gap between the ground and first excited states of the fermion-antifermion system cannot be interpreted as a threshold for production of a vector meson in a fermion-antifermion background. 

   One can also compute the time correlator of a gauge invariant, but spatially non-local operator, constructed by replacing
the Higgs field with an eigenstate of the covariant Laplacian operator, i.e.
\beq
         Q'_k(t) =   \sum_{\vx}  \zeta_1^\dg(\vx,t;U) U_k(\vx,t) \zeta_1(\vx + \hat{k},t;U) \ .
\label{Qp}
\eeq
The corresponding time correlation function is almost indistinguishable from $G(T)$ in \rf{GT}, and the mass is in agreement,
within error bars, with the vector boson mass extracted from the data in Fig.\ \ref{gmass}.  Note that if one used different
Laplacian eigenstates on the right and left side of $U_k$ in eq.\ \rf{Qp}, the time correlator would vanish, due to wild
fluctuations in the global phase of $\zeta_n(\vx;U)$.\footnote{One could construct gauge-invariant states containing two
Laplacian eigenstates and two vector bosons, which would be independent of the global phases.  The energy calculation would then require computation of a time correlation function of four vector boson operators, which I have not attempted.}

\begin{figure}[h!]   
\centerline{\includegraphics[scale=0.7]{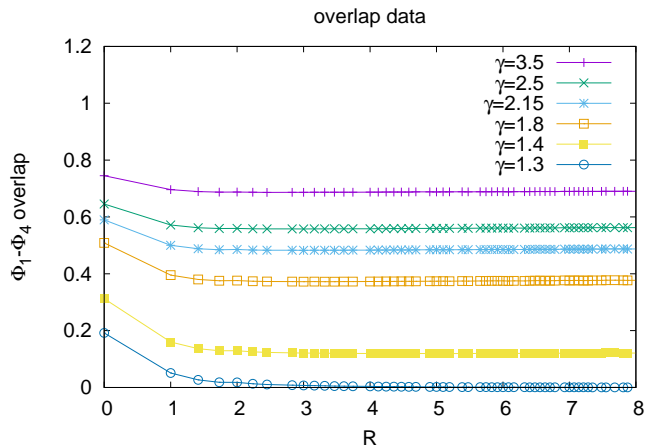}}
\caption{The overlap $o_{14}$  (from eq.\ \rf{o_mn}) vs.\ $R$ at fixed $\b=5.5$ and a selection of $\g$ values, where $o_{41}$ is the overlap between the normalized $\Phi_1$ and $\Phi_4$ states, which use pseudomatter and the Higgs field respectively to enforce gauge invariance. 
Note that $\g=1.3$ is either within or very near the confined phase, while all other $\g$ values are in the Higgs phase.}
\label{olapall}
\end{figure}

\begin{figure}[h!]
\subfigure[~]  % caption for subfigure a
{   
 \label{g215}
 \includegraphics[scale=0.6]{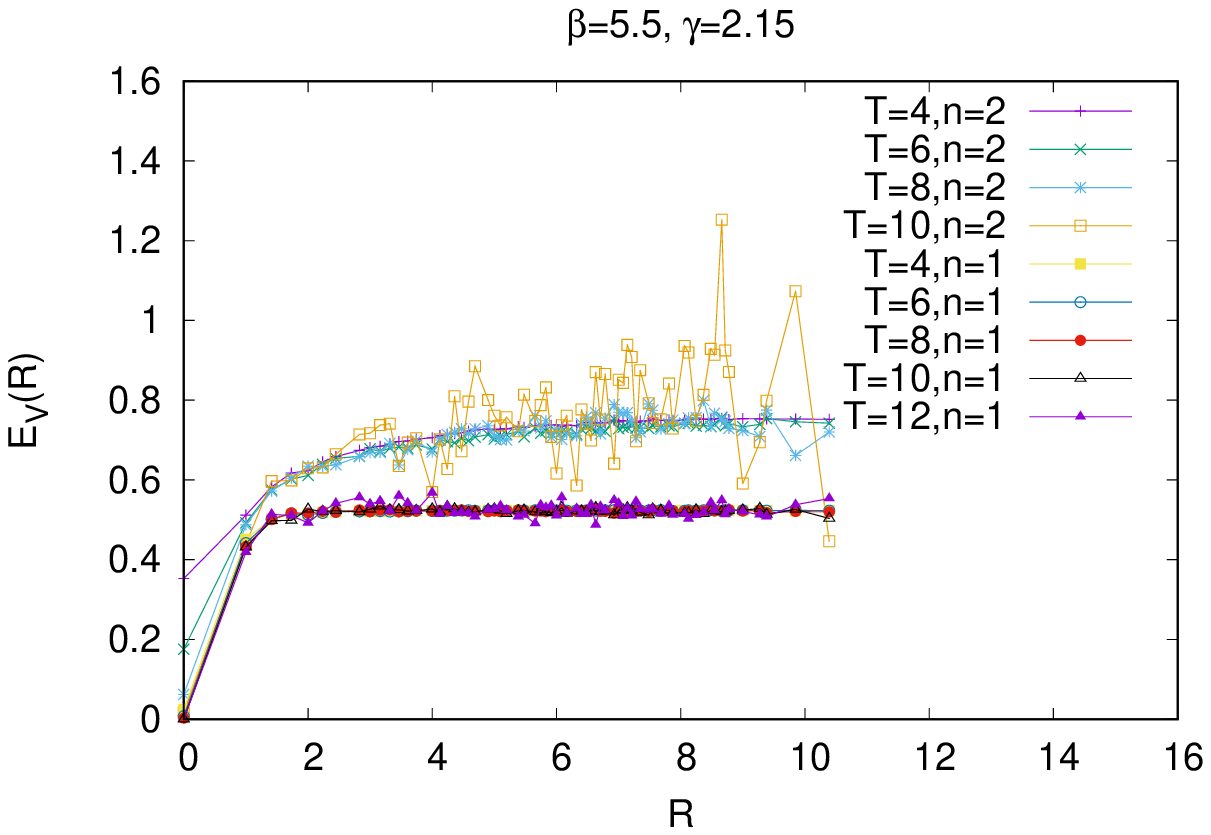}
}
\subfigure[~]
{
 \label{g180}
 \includegraphics[scale=0.6]{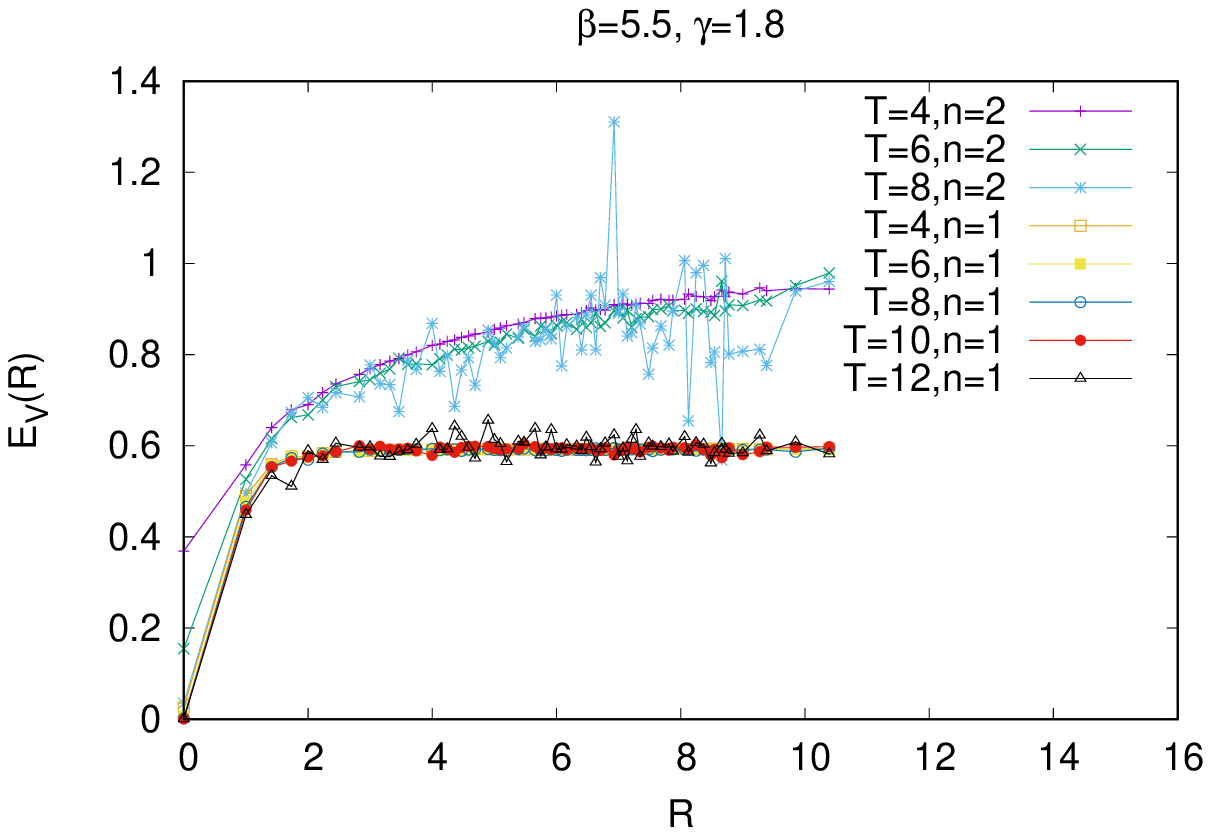}
}
\subfigure[~]
{
 \label{g140}
 \includegraphics[scale=0.6]{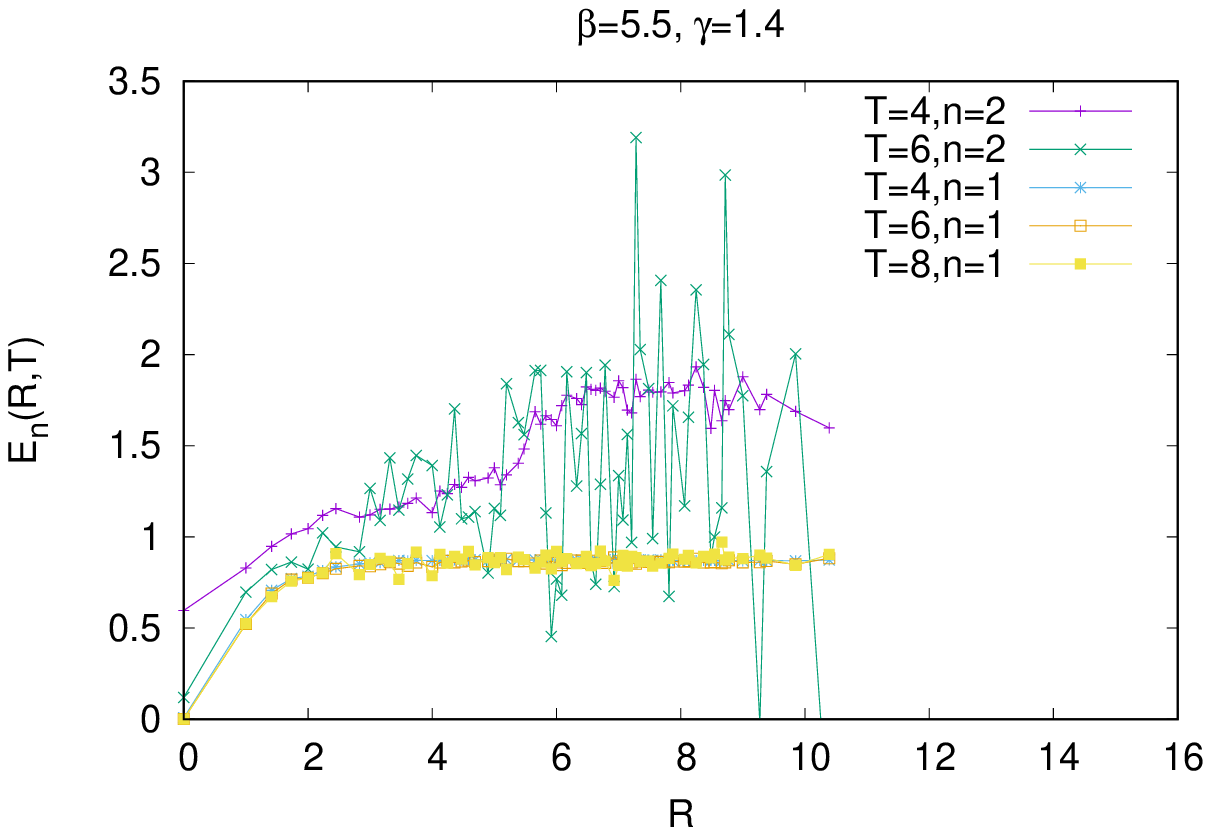}
}
\caption{Same as Fig.\ \ref{gap}, but at smaller $\g$ values, still at $\b=5.5$.  (a) $\g=2.15$; ~ (b) $\g=1.8$; (c) $\g=1.4$, which is just above the transition to the confined phase.  Note that the data for $n=2$, for data drawn from a fixed number (220) of
configurations, becomes noisy at smaller values of $T$, as $\g$ approaches the transition.  At $\g=1.4$ we cannot 
draw any conclusions about the large $T$ convergence of the $n=2$ data.}
\label{glow}
\end{figure}

    It is natural to ask what happens as $\g$ approaches the Higgs to confinement (spin glass to symmetric) transition at 
${\g \approx 1.35, ~ \b=5.5}$.   As
one might expect, the overlap $o_{14}$ between the normalized $\Phi_1$ and $\Phi_4$ states, built from pseudomatter and the Higgs field respectively, drops steadily as $\g$ is reduced, as seen in Fig.\ \ref{olapall}, and is consistent with zero at large $R$ at $\g=1.3$, which is either within or very near the confined phase. 

    As $\g$ is reduced below $\g=3.5$ in the Higgs phase at $\b=5.5$, both the ground and excited state energies gradually
rise, but an energy gap remains.  The data, for the fixed number of 220 lattices which were used at each $\g$, becomes noisier for $E_2(R,T)$ at small values of $T$ as $\g$ is reduced.  These tendencies are illustrated
in Fig.\ \ref{glow} for $E_n(R,T)$, again at $\b=5.5$ and $\g=2.15, 1.8, 1.4$.  The data for $E_2$ at $T=10$, for $\g=2.15$, and at $T=8$ for $\g=1.8$, has obviously quite a lot of statistical error, which presumably could be reduced by increasing statistics.  Still, the existence of an energy gap between the $n=1$ and $n=2$ levels appears to persist, at least at these lower $\g$ values.   
At a still lower value of $\g=1.4$, which is quite close to the Higgs/confinement transition, the data so far obtained for $E_2(R,T)$ is very noisy beyond $T=4$, and it is not possible to make any statement about convergence to a stable excitation.   \\

\section{Conclusions}

    I have constructed four gauge-invariant states for a static fermion-antifermion pair, at each fermion pair separation $R$, by combining the fermion operators with  Higgs and pseudomatter operators.  Working in the SU(3) gauge Higgs theory of eq.\ \rf{Sgh} at lattice couplings $\b=5.5$
and $\g=3.5$, it is found that one of these states ($\Psi_1$) converges rapidly, under Euclidean time evolution, to the ground state,
while the $\Psi_2$ state also rapidly converges, but to a state with an energy above the ground state.   Of course $\Psi_1$ and 
$\Psi_2$ are orthogonal by construction, but it appears that 
they remain almost orthogonal upon evolution in Euclidean time $T$.   What is significant is that the energy gap between the $n=1$ and $n=2$ states is nearly $T$ independent (for $T \ge 4$), and only weakly dependent on the fermion-antifermion separation $R$. The gap also seems to be small compared to the mass of the lowest lying particle
excitation, assuming (based on the results of \cite{Maas:2018xxu}) that the vector boson created by the operator
\rf{vecbos}  is the lightest particle.  This gap, below the vector meson threshold, indicates the existence of at least one gauge + Higgs field excitation of the fermion-antifermion system which cannot be readily interpreted as a fermion-antifermion ground state plus an additional particle.   It appears instead to be a stable excitation of the bosonic fields surrounding each of the elementary fermions.  Since the gap, judging from Fig.\ \ref{gap}, appears to have a finite limit at $R\ra \infty$, this is a physical excitation which is relevant to fermion-antifermion pairs at infinite separation.

    Of course these are only some first results indicating excitations of elementary fermions, and there are many open questions.  First, it would be helpful to have a more systematic examination of the particle spectrum
of the action \rf{Sgh}, to confirm that the vector boson associated with the operator \rf{vecbos} is in fact the lightest particle in the spectrum, and to map out the magnitude  of the excitation gap throughout the Higgs/spin glass phase of the phase diagram.  One would also like to generalize the action beyond special case of a unimodular Higgs field.
Secondly, we would like to know whether there are additional excitations of the fermion-antifermion system beyond the one found
here; perhaps this could be studied with a larger basis of $\Psi$ states, and much improved statistics.  Finally, it would be very interesting to know whether any of this is relevant to the electroweak theory, or to phenomenology beyond the Standard Model.  We reserve these questions for later investigation. \bigskip

\acknowledgments{This research is supported by the U.S.\ Department of Energy under Grant No.\ DE-SC0013682.}

\bibliography{sym3}

\end{document}